\documentclass{edp-conf}

\usepackage{graphicx}
\usepackage{natbib}
\usepackage{a4wide}
\usepackage{amssymb} 
\newcommand{\apj} {ApJ}

\newcommand{\aap} {A\&A}

\newcommand{\eqn} [1] {
\begin{equation} 
#1 
\end{equation}}

\newcommand{\eqna} [1] {
\begin{eqnarray} 
#1 
\end{eqnarray}}
\newcommand{\derivp} [2] {\frac {\partial #1 } {\partial #2} }

\newcommand{\ds}{\displaystyle}
\newcommand{\inv} {\frac {1}}

\newcommand{\deriv} [2] {\frac {d #1 } {d #2} }
\newcommand{\dprime}  {{\prime \prime}}

\begin{document}

\TitreGlobal{SF2A 2001}

%%-----------------------------
%%      the top matter
%%-----------------------------
\title{Stochastic excitation of stellar oscillations} 
\author{R. Samadi}\address{Astronomy Unit, Queen Mary, University of London, London E14NS, UK}
%\address{DESPA Observatoire de Paris-Meudon, 5 place J. Janssen, F-92195 Meudon, France}

%
\maketitle
\begin{abstract} 
Excitation of  solar oscillations is attribued to turbulent motions 
in the solar convective zone. 
It is also currently believed that oscillations of low massive stars 
($M <2~M_\odot$) -~which  possess an upper convective zone~-  are stochastically  excited  by  turbulent convection in their  outer layers.
%Providing that accurate measurements of the oscillation amplitudes and  damping rates are available it is possible to evaluate the power  injected into the modes and thus - by comparison with the observations - to constrain current theories of stochastic excitation of stellar oscillations.

A recent theoretical work \citep{Samadi00I,Samadi00II} supplements and
reinforces 
this theory. This allows the use of any available model of turbulence
%process was generalized to a global description of the turbulent state of their convective zone 
%The comparison between observation and theory, thus generalized, will allow to
%better know the turbulent spectrum of stars, and this in particular thanks to
%the COROT mission. The present report  summarizes  these results 
and emphasizes some recent unsolved problems which are brought up by 
these new theoretical developments. 

%We compare  our amplitudes computation with recent observation of Procyon~A and $\alpha$~Cen~A. 
%Finally we summarize results of a systematic computation of the oscillation power spectrum of  solar-like oscillating stars with different effective temperatures and luminosities. 

 \end{abstract}

%
%%-----------------------------
%%      your text
%%-----------------------------
\section{Introduction}

% The  amplitudes of solar-like oscillations    result from a balance between excitation and damping. 
%However our understanding of the excitation and damping processes remains still poor.
%The accurate measurement of the amplitudes and the growth rates $\eta$ - which will be possible in particular with the  COROT  mission -  will  make it possible to constrain the theory of the excitation and damping. 

%It was suggested more than  thirty years ago that the  stochastic excitation process   by  turbulent convection  in the Sun is at the origin of the low amplitudes of the   solar oscillations. % (about $10^{-6}$ in term of relative fluctuation of luminosity). 
The excitation of  the solar oscillations results from the action of the
Reynolds 
tensor  and of the turbulent fluctuations of entropy; the latter comes from a heat exchange between the oscillations and the turbulent elements. 
The nature and the properties of the source term of excitation due to the
Reynolds tensor  are well established  \citep{GK77}, 
although the evaluation of this term remained crudely approximated. 
On the other hand,  the source of excitation which takes its origin in  
the fluctuations  of the entropy, had given rise to major 
controversies \citep[e.g.][GMK hereafter]{GMK94}.
% existed and we isolated an inconsistence in his evaluation by some authors.

In \citet[Paper~I hereafter]{Samadi00I},  
inconsistencies in the available theories were removed 
which led to a formulation which
 generalizes results from previous works,  and 
is built so as to enable consistent investigations of  
various  possible spatial and temporal  spectra of stellar turbulent convection.
 It also shows that the  actual entropy  source term results from the advection of 
the turbulent fluctuations of entropy by the turbulent motions.
These  results are  summarized in Section~\ref{sec:Theory of the stochastic excitation} and Section~\ref{sec:Results in the Solar case}.

The unavoidable free parameters introduced in the formulation
 are adjusted in order to obtain the best  fit to the solar seismic observations. 
In a second step, the comparison between the computed amplitudes  and  the observations 
as well as the use of  theoretical  arguments allow  us  to determine 
the ingredients in the theory which are still defective. 
Section~4  then lists some unsolved problems which are brought up 
by our improvment of the  theoretical approach in Paper~I.
Final comments conclude  about elements of the theory which still remain to be
 developed.

%The free parameters introduced in the theory are first  adjusted to obtain the best fit to the solar observations.
%We then consider  stellar models o Procyon~A and $\alpha$~Cen A for which we compute the oscillation amplitude spectrum. Results are next compared with the recent observations.

%%%%%%%%%%%%%%%%%%%%%%%%%%%%%%%%%%%%%%%%%%%%%
\section{Theory of the stochastic excitation}
\label{sec:Theory of the stochastic excitation}

The stochastic excitation mechanism results 
from a forcing of the stellar material by the turbulent motions 
and by a heat exchange due to the turbulent fluctuations of the entropy;
in other words, the acoustic power generated by the turbulence 
 excites resonant modes of the 
stellar cavity (oscillations). 
%We summarize below the theoretical approach adopted in Paper~I 
%to model the way the acoustic modes are exicted.

In the presence of turbulence one shows that the radial 
oscillation velocity $\vec v_{osc}$ obeys the  inhomogeneous 
wave equation (see Paper~I for details):
\eqna{
\rho_0 \left ( \derivp  { ^2 } {t^2}  -  \vec L  \right )
\left [  \vec v_{osc}  \right ] +
\vec {\cal D} (\vec v_{osc})  & = & 
 \derivp{}{t}
\left (  \vec f_{t}  + \vec \nabla  h_t   \right ) 
\label{eqn:inhomogeneous_wave}
}
with
\eqna{
 \vec f_{t} = -\vec \nabla :  (\rho_0\vec {u} \vec{u} )
\label{eqn:f_t} \\
\derivp{}{t} \vec \nabla h_t 
=  -  \vec \nabla \left  ( \right . \alpha_s \deriv{}{t} \delta s_t - 
\vec \nabla. (\alpha_s \vec u \, s_t ) +
 s_t \vec u . \vec \nabla \alpha_s \left . \right )
\label{eqn:h_t}
}
the turbulent Reynolds stress and the source term due to the turbulent 
entropy fluctuations respectively. The expression for the wave
operator $ \vec L $ is given in Paper~I.
In Eqs.(\ref{eqn:inhomogeneous_wave} , \ref{eqn:f_t}, \ref{eqn:h_t}),  
$\rho$, $s$, $c_s$,  $\vec g$ and $\vec u$ respectively 
denote density, entropy, sound speed, gravitational acceleration 
and velocity of the turbulent elements;  the subscript 0 refers to average mean quantities.

The first term in the RHS of Eq.(\ref{eqn:h_t}) is due to the Lagrangian
entropy fluctuation $\delta s_t$. % which  has been considered by GMK and B92. 
The last two  terms are due to the buoyancy force associated with the Eulerian 
entropy fluctuations $s_t$. These  terms are found to contribute  to the excitation
as much as  the Reynolds source term. Other  source terms
 in the RHS of  Eq.(\ref{eqn:inhomogeneous_wave}) are
 found negligible in Paper~I and are discarded here.

The operator $\vec {\cal D}$ in Eq.(\ref{eqn:inhomogeneous_wave})  
is responsible for generating both  a dynamical damping and a 
shift of the oscillation frequency due to the action of 
the turbulent elements. 
Contribution of the operator $\vec {\cal D}$ to the damping
 is expected to be small compared to those of the other damping processes. 
and is  assumed to be included 
in the global damping rate $\eta$ which  takes into account all damping processes.

%______________________________
%\subsection{Free oscillations}

Assuming  no turbulence ($\vec u = 0$) 
the velocity field is simply related to the pulsational radial displacement as $\ds{
\vec{v} = \deriv {}{t}   \vec \xi_r \left (\vec{r},t  \right )} $
where $\vec \xi_r (\vec{r},t)$ is the undriven pulsational radial
displacement in the absence of turbulence.
 Eigensolutions of Eq.(\ref{eqn:inhomogeneous_wave}) complemented with 
boundary conditions, can be written,  
in the absence of turbulence, in terms of  the   displacement 
$\vec {\xi}_r (\vec{r},t) = \vec \xi_r(\vec r) \, e^{-i\omega_0 t}$ 
where  $\omega_0$ is the oscillation frequency 
and $\vec \xi_r (\vec r)$  is the adiabatic (real) displacement  eigenvector 
which satisfies $\ds{ \vec L(\vec \xi_r (\vec r))  =  - ~ \omega_0^2 \vec \xi_r (\vec r)
} $.

%______________________________
%\subsection{Oscillations excited by turbulence}

When turbulence is present, the pulsational displacement  and  velocity are 
 written   in terms of the above adiabatic 
solution $\vec {\xi}_r (\vec{r},t)$  and an instantaneous
amplitude   $A(t)$.
Accordingly 
\eqna{
\delta \vec r_{osc} 
=  {1\over 2} \left (A(t) \,  \vec {\xi}_r (\vec r) \,  e^{-i \omega_0 t } +cc \right )  & \textrm{and} &\vec v_{osc} =  {1\over 2} 
(-i \omega_0 \,  A(t) \,  \vec {\xi}_r (\vec r) \,  e^{-i \omega_0 t } +cc)
\label{eqn:vosc}
}
where cc means complex conjugate. 
The mode energy, averaged over a time scale smaller than the mode lifetime $\eta^{-1}$ and larger than the eddies time scale,  is given by
\eqn{
E=  \int_0^{M} dm ~\langle   v_{osc}^2  \rangle 
=   \inv {2} ~\langle |  A | ^2 \rangle  \; I {\omega_0}^2
} 
where $\ds{I \equiv   \int_V  \rho_0 \, d^3 x\,  (\vec \xi_r^*  . \vec \xi)_r }$
is the mode inertia and  $\langle  |A| ^2 \rangle $ is the mean square
 amplitude.
Finally a general expression for the power, $P$, going into each mode 
is established in Paper~I: 
\eqna{
P\equiv \deriv{E}{t}  =  \eta\,  {\langle A^2 \rangle}\,I\,\omega_0^2 
\label{eqn:power} & \textrm{with} & 
\left < \mid A \mid ^2 \right >  = \frac{1}{8 \eta (\omega_0 I)^2}  
\left ( C_R^2 + C_S^2  \right )
\label{eqn:A2_0}
}
where $C_R^2$ and $C_S^2$ are the turbulent Reynolds stress and 
the entropy contributions respectively given by:
\eqna{
C_R^2 & =& \int  d^3x_0   \,  \int_{-\infty}^{+\infty} d\tau \, e^{-i\omega_0 \tau}  \int d^3r\left < \, 
\left (\rho_0 u_j u_i  \nabla^j  \xi_r^i \right )_1  
\left (\rho_0 u_j u_i \nabla^j  \xi_r^i \right )_2    
\right >  
\label{eqn:C2R} \\
C_S^2 & =  &   \int  d^3x_0   \,  \int_{-\infty}^{+\infty} d\tau \, e^{-i\omega_0 \tau}  
\int d^3r  \left < \, 
 \left ( h_t \, \vec \nabla \, .  \, \vec \xi_r      \right)_1  
\left (  h_t \,  \vec \nabla \,  . \, \vec \xi_r    \right )_2  \,
 \right >  
\label{eqn:C2S}
}
where $\vec x_0$ is  the average position where the 
power is evaluated whereas $\vec r $ and $\tau $ are
 related to the local turbulence.
Subscripts   1 and 2 refer to the spatial and  
temporal positions $[ \vec x_0-\frac{\vec r}{2}, - \frac{\tau} {2}]$  
and $ [\vec x_0+\frac{\vec r}{2}, \frac{\tau} {2}]$ respectively. 
%We adopt the Einstein convention of summation upon repeated indices. 

%------------------------------------------------
\subsection{Reynolds stress contribution}

It is shown that the terms $(\rho_0 \nabla_1^i \xi_r^j ) $ 
and $(\rho_0 \nabla_2^i \xi_r^j)$ in Eq.(\ref{eqn:C2R})  
do not change  over the  length scale of the eddies. 
Consequently,  in Eq.(\ref{eqn:C2R}), 
integrations over $\tau$ and $r$ only involve
 the phase term  $e^{-i\omega_0 \tau} $ 
and the fourth-order velocity correlations  
$  \langle u_i^\prime u_j^\prime u_l^\dprime u_m^\dprime \rangle$ where $\ds{
\vec{u}^\prime = \vec{u}(\vec x_0-{\vec r}/{2},-{\tau}/{2} ) }$ 
and $\ds{\vec{u}^\dprime = \vec{u}(\vec x_0+{\vec r}/{2},+{\tau}/{2} )}$.

The  Quasi-Normal Approximation  
\cite[Chap VII-2, QNA hereafter]{Lesieur97} is adopted 
so that  the fourth-order velocity correlations  can be written as 
 a product of second-order  velocity correlations  
$  \langle u_i^\prime u_j^\dprime  \rangle$.
The properties of $\langle u_i^\prime u^\dprime_j \rangle $ are
 well known in the Fourier domain $(\vec k,\omega)$
 where $\vec k$ and $\omega$ are the wavenumber and
 the frequency associated with a turbulent element.
The  second-order velocity correlations  can be expressed
 formally in term of  $ E( k,\omega) $ the turbulent kinetic energy spectrum.
Following  \citet{Stein67}, the velocity
 energy spectrum $ E(k,\omega) $ is written as
\eqna{
E( k,\omega) =E(  k) \, \chi_k(\omega) 
\label{eqn:ek_chik}
} 
where $\chi_k(\omega)$ is a  frequency-dependent factor which can be related 
to the time correlation of the eddies in the frequency space.
Finally  a general expression for $C_R^2$ involving $E(k)$ and $\chi_k(\omega)$
is given in Paper~I:
\eqna{
C_R^2 & = &  4 \pi^{3} \,  {\cal G}  \int_{0}^{M} dm   \,
\rho_0 \left (\deriv { \xi_r} {r} \right )^2 
\int_0^\infty dk \, \frac {E^2(k)} { k^2}  \chi_k ( \omega_0 )
\label{eqn:C2R_3}
}
where   ${\cal G}$  is a constant anisotropic factor. % similar to the anisotropic factor introduced by \citet{Gough77}.

%------------------------------------------------
\subsection{Contribution of entropy fluctuations}

We assume in Paper~I that $\delta s_t$ and $s_t$ in Eq.(\ref{eqn:h_t}) 
act as  passive scalars. Let $E_s(k,\omega)$ be the 
spectrum of the scalar correlation product 
$\left  < s_t^\prime s_t^\dprime \right >$. 
It is also assumed that  the frequency-dependent component 
of  $\left  < s_t^\prime s_t^\dprime \right >$ is 
the same than those of the velocity field correlation 
product $\langle u_i^\prime u^\dprime_j \rangle $ and
 that  $E_s(k,\omega)$ can be decomposed as $E(k,\omega)$ in Eq.(\ref{eqn:ek_chik}). 
%The theory of the turbulence allows one to related $E_s(k)$ to $E(k)$ (see Paper~II).

As a consequence of the above assumptions,  it is then demonstrated 
that the correlation product of the Lagrangian
 entropy fluctuations $\langle \delta s_t^\prime \, \delta s_t^\dprime \rangle$
vanishes after  integration over $\vec r$ in Eq.(\ref{eqn:C2S}).
This result may  be explained as follows: turbulence  and oscillation  
are  coupled through the phase term  $e^{-i \omega_0 \tau}$ and through the 
turbulent time spectrum $\chi_k(\omega)$. 
Therefore this coupling occurs at frequencies
 close to the oscillation frequency $\omega_0$.
It then follows that the coupling between turbulence and  oscillation involves 
eddies of wavenumber $k \gg k_{osc} $ where $k_{osc}$ is 
the wavenumber of the mode and $k$ that of an eddy. 
On the other hand the spatial component 
of  $\langle \delta s_t^\prime\,  \delta  s_t^\dprime
 \rangle$ in the Fourier space favors eddies with the 
largest size ($k \rightarrow 0$).
These two opposite effects  clearly are incompatible and lead to a vanishing 
 contribution for  the Lagrangian entropy fluctuation.

This does not happen for the contribution of the Reynolds  source term 
which involves  the fourth-order velocity correlation product. 
According to the QNA this term  can be decomposed in term of a product of two  
second-order velocity correlations.	
Coupling with the oscillation then becomes  non-linear and leads to
 an effective non zero contribution. Thus {\it only non-linear 
terms, with respect to the fluctuations, can contribute to mode excitation while linear terms  do not}. 
This may be considered as a general result.

With an analogous procedure than for  
the Reynolds stress contribution, we show that 
the entropy contribution can 
 be expressed as: 
\begin{eqnarray}
C_S^2 &=&  \frac{ 4 \pi^3 \,{\cal H} } {\omega_0^2} \int d^3 x_0 
 \left ( \alpha_s \deriv{\xi_r}{r}   \right ) ^2   g_r  
    \, \int  dk \,  \frac { E_s(k)  E(k) } { k^2 } 
\int_{-\infty}^{+\infty} 
d\omega \, \chi_k(\omega_0+ \omega) \chi_k(\omega) 	
\label{eqn:C2S_7} 
\end{eqnarray}
with $g_r(\vec \xi_r,r)$ a function which expression is given in Paper~I
and  ${\cal H}$  a constant anisotropic factor similar to  ${\cal G}$.

%%%%%%%%%%%%%%%%%%%%%%%%%%%%%%%%%%
\section{Results in the Solar case}
\label{sec:Results in the Solar case}

%--------------------------------------------
\subsection{Models for the solar turbulence}
\label{Models for the stellar turbulence}

On one hand,  the turbulence theory tells us   that $E(k)$ 
follows the Kolmogorov spectrum as $E(k) \propto k^{p}$ 
with the slope $p=-5/3$. 
On the other hand, observations of the solar granulation 
allow one to determine the turbulent kinetic spectrum $E(k)$ of the Sun.

Several  kinetic turbulent spectra 
have been   suggested by different  observations of the solar granulation:
the  ``Raised Kolmogorov Spectrum `` (RKS hereafter) , the ``Nesis Kolmogorov Spectrum'' (NKS hereafter) and the ``Broad Kolmogorov Spectrum'' (BKS hereafter).  
These spectra (Figure \ref{fig:adjust}) obey the Kolmogorov law 
for $k \geq k_0$ where $k_0$ is the wavenumber at which the turbulent cascade
begins. 
They differ in the injection region ($k < k_0$) where they are 
characterized by different values for the slope $p$.

The wavenumber $k_0$ is unknown but can be related  to the mixing length
 $\Lambda$ as $k_0=2\pi \, / \, (\beta \Lambda)$ where  $\beta$ is  a 
free parameter introduced for the arbitrariness of such definition (Paper~I).
Moreover the definition of the eddy correlation time , which
 enters the description of the
 turbulent excitation,  is somewhat  
arbitrary and is therefore  gauged by introducing an additional free parameter
 $\lambda$. 
%We show that the oscillation power computed in the case of solar oscillations is very sensitive to the values of the free parameters $\lambda$,  $\beta$  (Paper~II). 

\subsection{Oscillation power}
\label{Oscillation power}

The power injected into the solar oscillations 
is related to the rms value $v_s$ of the surface velocity  as
$ v_s^2  (\omega_0) = \xi_r^2(r_s)  \;  P (\omega_0)\, / \, 2 \eta I $
where  $r_s$ is the radius at which oscillations are measured.
Observations of the solar oscillations provide $v_s$ and $\eta$ such that 
$P(\omega_0)$ can be evaluated from pbservations. On the theoretical side, 
 the power $P(\omega_0)$  is computed according to Eqs(\ref{eqn:power}, \ref{eqn:A2_0}, \ref{eqn:C2R_3}, \ref{eqn:C2S_7}) for  a calibrated solar model. 
The physical ingredients of the model are detailed in \citet[Paper~II
hereafter]{Samadi00II}.
As was  done in Paper~II, the free 
parameters $(\beta,\lambda)$ are adjusted in order 
to obtain the best  fit to the solar seismic observations 
by \citet{Libbrecht88}
% However, in Paper~II, the free parameters 
%were adjusted in order to minimize the shift in frequency between 
%the computed amplitudes spectrum and the observed one.
%The high frequency modes ($\nu \gtrsim 3.5$~mHz) 
%are mainly excited by small size turbulent elements. 

Theoretical arguments support  the fact that 
the current theory of stochastic excitation 
 is less justified for the low frequency modes than for the 
high frequency ones (see below section~\ref{Static 
and dynamic properties of the turbulent medium}). Therefore in the following
$\beta$ and $\lambda$ are  adjusted  in order to reproduce,  as best as 
possible, the frequency dependency of $v_s(\omega_0)$ 
at high frequencies ($\nu \gtrsim 3.5$~mHz).

\begin{figure}[ht]
\begin{center}
\resizebox{15cm}{!}{\includegraphics{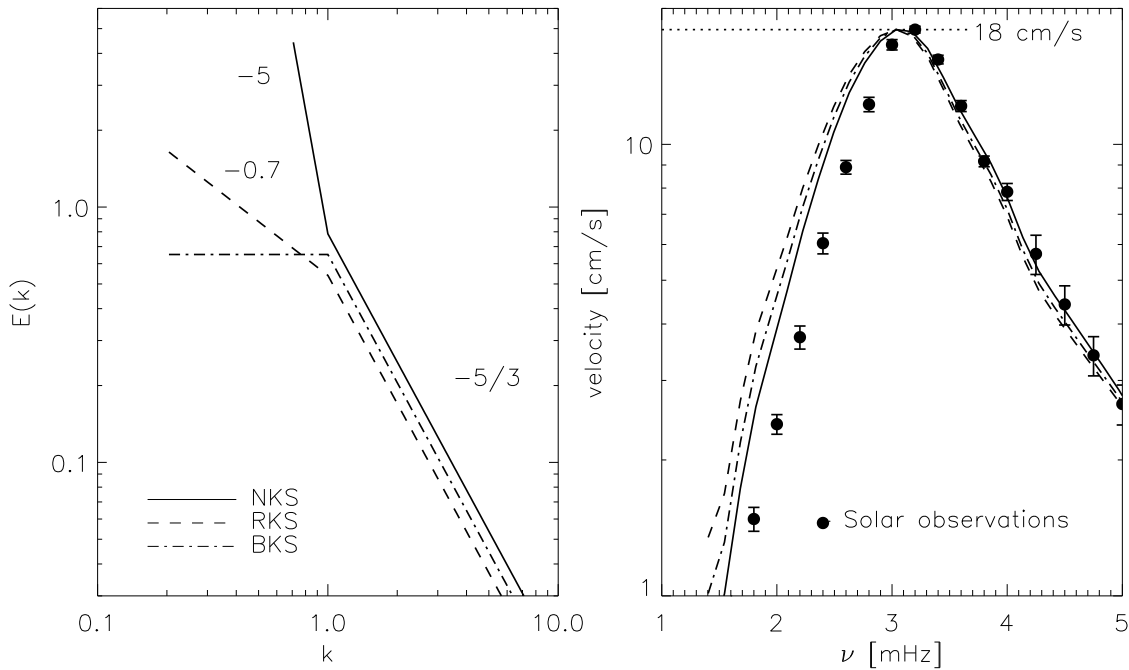}}
\caption{{\bf Left panel:}  Kinetic turbulent spectra versus wavenumber $k$.
{\bf Right panel:} Computed surface velocity $v_s$ assuming the turbulent spectra plotted on the left panel. $\lambda$ and $\beta$ values result from fitting the computed $v_s$ to the solar seismic observations by \citet{Libbrecht88} using the observed damping rate $\eta$. }
\label{fig:adjust}
\end{center}
\end{figure}

Results of the fitting are shown in Fig. \ref{fig:adjust}.
As a consequence  of the  adjustement,  all the kinetic turbulent spectra fit well the solar observations at
 high frequency ($\nu \gtrsim 3.5$~mHz) 
while the main differences are observed at low frequency. The
overall best agreement is obtained with the NKS.

Moreover we find that the entropy source term  significantly contributes to 
the excitation process of solar-like oscillations by turbulent convection,
in agreement with the results of GMK and simulations by  \citet{Stein91}.

%%%%%%%%%%%%%%%%%%%%%%%%%%%%%%%%
\section{Some open issues}

%----------------------------------------------------------------
\subsection{Static and dynamic properties of the turbulent medium}
\label{Static and dynamic properties of the turbulent medium}

 As stressed by \citet{Rieutord00}, differences at low wavenumber 
between the kinetic spectra obtained from different observations 
of the solar granulation are a consequence of uncontrolled 
data-averaging procedures. Mean average (static) properties of
 the kinetic spectrum at low wavenumber (mesogranulation) are 
thus not yet well represented by the observations of the solar surface. 
In the same way, \citet{Nordlund97} demonstrated with a 
numerical simulation that observations of the solar granulation 
cannot provide a meaningful determination of the turbulent properties.
Therefore, the mean average (static) properties of the different 
turbulent spectra -  suggested by the current observations 
of the solar granulation - are not yet well established.

On the theoretical side,  the derivation of Eq.~(\ref{eqn:C2R_3}) 
and Eq.(\ref{eqn:C2S_7}) uses the decomposition of Eq.(\ref{eqn:ek_chik}). 
$E(k)$  depicts the static properties of the turbulent medium 
whereas $ E(k,\omega) $ characterizes  the dynamic properties.
In Paper~II different forms for  $\chi_k(\omega)$ -~ 
which were suggested by \citet{Musielak94}~- were tested. 
% We have thus considered the Gaussian form the modified gaussian form and an exponential form (see Paper~II ).
From the comparison with the solar seismic data we cannot clearly 
discriminate between the different forms and 
% modified gaussian form and the gaussian form and  
we finally adopted the gaussian form (Paper~II).

However it is suggested in \citet{Samadi00Phd} that the 
decomposition of Eq.(\ref{eqn:ek_chik}) is not valid 
at small wavenumber (in the injection region,  i.e. for $k \lesssim k_0$).
Indeed the high frequency modes involve small size turbulent elements ($k > k_0$).
At this scale (the inertial range) the lifetime $\tau_k$ of the turbulent 
elements are shorter than the  timescale $\tau_\Lambda$  
at which energy is injected into the turbulent cascade.
Therefore at small scales  injection of energy into the cascade
 appears stationary and the  dynamic properties of $E( k,\omega)$ 
are only fixed by the lifetime $\tau_k$ of the eddies. 
The line-width of the function $\chi_k(\omega)$ provides  a 
statistical measure of $\tau_k$.
As a consequence, the decomposition of  Eq.(\ref{eqn:ek_chik}) 
should be valid for $k \gtrsim k_0$.

In contrast at large scales,  ( $k \lesssim k_0$), 
the lifetime of the turbulent elements is of the same order 
than $\tau_\Lambda$.
Therefore the energy transfer between large and small elements 
is no longer  stationary  and consequently 
the decompostion of  Eq.(\ref{eqn:ek_chik})  is probably no longer valid.

This statement explains  the large changes observed 
at low frequency between the different spectra (Fig.~\ref{fig:adjust}). 
Indeed the  low frequency modes involve the largest eddies and 
the changes at low frequency are therefore due to the differences
 in the static properties of the turbulent spectra 
in the injection region (see Fig.~\ref{fig:adjust}).
The best agreement is obtained with the NKS which however 
appears as the less realistic spectrum.
In contrast the more plausible spectra (RKS and BKS) 
significantly overestimate  the amplitudes at low frequency.
These results prove that the dynamic properties of $E( k,\omega)$ is 
not correctly modeled through the decomposition of  Eq.(\ref{eqn:ek_chik}) 
in the range $k \lesssim k_0$.
Moreover,  the decomposition of Eq.(\ref{eqn:ek_chik}) is also used for 
 the turbulent spectrum $E_s(k,\omega)$ of the entropy fluctuations. 
Therefore the above conclusions  also apply to $E_s(k,\omega)$.

%-----------------------------------------------------
\subsection{Excitation of the non-radial oscillations}

The  present theory of stochastic excitation only concerns  the radial $p$-modes.
It is shown in \citet{Samadi00Phd}  that the stratification does not 
affect the  transfer of acoutic power into the radial modes. 
%Moreover it has been proved that the radial oscillations and the contributive eddies are well decoupled each other \citep{Samadi00I}.
%These two features permit us to separate the turbulent source terms from the startification and the mode eigenfunction leading to an analytical expression for the power injected into the the radial modes.
%For radial modes, the stratification does not affect the  emission of power because most of the contributive eddies are perpendicular to the modes and as a consequence the contributive eddies lie in the horizontal planes.

The high $\ell$ degree modes propagate in both radial and 
horizontal directions. The acoustic power injected into 
high $\ell$ degree modes is therefore more sensitive to the stratification.

The current theory is mainly based on the approximation that 
the stratification and the sources of excitation are well decoupled.
\citet{Stein67} showed that the acoustic emission arising 
from the stratification is negligible compared to the Reynolds stress emission.
However some high $\ell$ degree modes propagate mainly in the 
horizontal direction in the excitation region. For such modes it 
is important to evaluate the contribution  of the stratification 
to the acoutic power emission.
%This aspect of the theory should be investigated both theoretically and with the help of 3-D simulations.

%%%%%%%%%%%%%%%%%%%%%
\section{Conclusion}

The present paper summarizes the theory  of 
solar oscillation excitation at it is exposed in \citet{Samadi00I}. 
In particular  assumptions and approximations
 which were used are recalled here with some details. 
Open issues in this field are also discussed 
such as  the way the turbulence is modeled and the case of non-radial modes:

%Theoretical arguments support the fact that the current model of 
%turbulence adopted in the theory is not valid  at 
%large scale length (meso-granulation).
%Moreover current observations of the solar granulation 
%do not permit us to establish with confidence 
%the mean average (static) properties 
%and the dynamic one of the the turbulent medium.
The way the turbulence is currently modeled at large scale 
length introduces  large uncertainties in the 
computation of the  power injected into the low frequency modes 
($\nu \lesssim 3.3$~mHz in the solar case).
A more  robust modeling of the large scales is thus needed in 
order to extend to  low frequencies the validity of the current theory. 
This can be achieved with the use of 3D simulations of the solar 
convection zone (work in progress).

Current theories only concern the excitation of radial modes.
For non-radial modes effect of the stratification upon the 
excitation may be important whereas it is negligible for radial ones.
Therefore effects of the stratification to the power emission 
in the high $\ell$ degree modes, as well as the $\ell$ dependency 
of the current theory of stochastic excitation, should be addressed (work in progress).

%\bibliography{/home/reza/predict/redac/biblio}
\bibliographystyle{astron}

\end{document}